\documentclass[twocolumn,showpacs,preprintnumbers,amsmath,amssymb]{revtex4}

\usepackage{graphicx}
\usepackage{dcolumn}
\usepackage{bm}

\begin{document}

\preprint{APS/123-QED}

\title{Elastic properties of the Non-Fermi liquid metal CeRu$_{4}$Sb$_{12}$ and the Dense Kondo semiconductor CeOs$_{4}$Sb$_{12}$\\}

\author{Yoshiki Nakanishi}
\email{yoshiki@iwate-u.ac.jp}
\author{ Tomoyuki Kumagai }
\author{ Masafumi Oikawa}
\author{ Tomoaki Tanizawa}
\author{Masahito Yoshizawa}

\affiliation{%
Graduate School of Frontier Materials Function Engineering, Iwate University Morioka 020-8551, Japan
}%

\author{Hitoshi Sugawara }
\email {Present address: Faculty of Integrated Arts and Sciences, The University of Tokushima Tokushima 770-8502, Japan }
 \author{Hideyuki Sato}
\affiliation{Department of Physics, Tokyo Metropolitan University, Hachioji 192-0397, Japan 
}%

\date{\today}

\begin{abstract}
	We have investigated the elastic properties of the Ce-based filled skutterudite antimonides CeRu$_{4}$Sb$_{12}$ and CeOs$_{4}$Sb$_{12}$ by means of ultrasonic measurements. CeRu$_{4}$Sb$_{12}$ shows a slight increase around 130 K in the temperature dependence of the elastic constants $C$$_{11}$,  ($C$$_{11}$-$C$$_{12}$)/2 and $C$$_{44}$. No apparent softening toward low temperature due to a quadrupolar response of the 4$f$-electronic ground state of the Ce ion was observed at low temperatures. In contrast CeOs$_{4}$Sb$_{12}$ shows a pronounced elastic softening toward low temperature in the longitudinal $C$$_{11}$ as a function of temperature ($T$) below about 15 K, while a slight elastic softening was observed in the transverse $C$$_{44}$ below about 1.5 K. Furthermore, CeOs$_{4}$Sb$_{12}$ shows a steep decrease around a phase transition temperature of 0.9 K in both $C$$_{11}$ and$ C$$_{44}$. The elastic softening observed in $C$$_{11}$ below about 15 K cannot be explained reasonably only by the crystalline electric field effect. It is most likely to be responsible for the coupling between the elastic strain and the quasiparticle band with a small energy gap in the vicinity of Fermi level. The elastic properties and the 4$f$ ground state of Ce ions in CeRu$_{4}$Sb$_{12}$ and CeOs$_{4}$Sb$_{12}$ are discussed from the viewpoint of the crystalline electric field effect and the band structure in the vicinity of Fermi level.
\end{abstract}

\maketitle

\section{\label{sec:level1}Introduction}

Filled skutterudite compounds, RETr$_{4}$X$_{12}$ (RE: rare earth; Tr=Fe, Ru, Os; X: pnictogen) exhibit a wide range of electrical and magnetic properties, mainly due to the 4$f$-electronic state of RE.[1] In this family, semiconducting behavior is seen mostly in Ce- and U- compounds.[2-4] The energy gap formation in dense Kondo systems is one of the most interesting subjects in strongly correlated electron systems. The strong electric correlation effect of $f$-electrons plays a crucial role in the systems. In the Ce-compounds of CeFe$_{4}$P$_{12}$, CeFe$_{4}$As$_{12}$, CeRu$_{4}$P$_{12}$ and CeOs$_{4}$P$_{12}$ the energy gap has been estimated to be of 110 meV, 10 meV, 86 meV and 34 meV, respectively. [3,5, 6] The energy gap is approximately proportional to the lattice constant in Ce-based filled skutterudites. More interestingly, the non-Fermi liquid (NFL) behavior appears by the disappearance of the energy gap in the vicinity of Fermi level at low temperatures, which is realized in CeRu$_{4}$Sb$_{12}$.[7-9] Furthermore, CeRu$_{4}$Sb$_{12}$ exhibits intermediate valence behavior in the magnetic susceptibility and electrical resistivitiy in which an anomalous temperature dependence appears below about 100 K. The electronic contributions to the specific heat is rather large with 380 mJ/mol K$^{2}$.[7] Recent optical and ultrahigh resolution photoemission studies found that CeRu$_{4}$Sb$_{12}$ exhibits a charge gap feature below 70 K with the energy gap of $\Delta$ = 47.1 meV.[10] A further interesting behavior: NFL appears below about 1K in the specific heat with the unusual ln$T$ dependence and below 5 K in the resistivity with a $T$$^{1.65}$ dependence.[7] 

Recently, Bauer $et$ $al$., have found that the resistivity of CeOs$_{4}$Sb$_{12}$ exhibits a semiconductor-like behavior at low temperatures with energy gap of 5-15 K by the use of an activated conduction law in the temperature range 25 K $<$ $T$ $<$ 50 K.[11] They also estimated the Kondo temperature $T$$_{K}$ to be about 90 K. The relatively large specific heat coefficient $\gamma$=$C$/$T$ of 96 mJ/mol K$^{2}$ indicates the formation of Heavy Fermion system at low temperatures. These experimental results indicate that the dense Kondo system with the energy gap at the Fermi level is realized in CeOs$_{4}$Sb$_{12}$. The magnetic susceptibility was explained by the crystalline electric field (CEF) effect with the energy splitting of $\Delta$ = 327 K between the $\Gamma$$_{7}$ doublet ground state and the $\Gamma$$_{8}$ quartet excited state. It also suggested that the Ce ions are almost trivalent in CeOs$_{4}$Sb$_{12}$. The lattice parameters of REOs$_{4}$Sb$_{12}$ follow the lanthanide contraction at room temperature, at least.[12] The well-localized nature of Ce ions expected by the CEF effect is not consistent with the character of a dense Kondo system with an energy gap. Furthermore, CeOs$_{4}$Sb$_{12}$ shows a pronounced anomaly at 0.9 K in the specific heat, indicating an intrinsic phase transition.[13-14] However, the estimated entropy is extremely small amount of 2$\%$ of $R$ln2 which is expected by the $\Gamma$$_{7}$ doublet ground state. Thus, one can expect that this phase transition is not ascribable to the well-localized 4$f$ electronic state derived from the Ce ion. Interestingly, this transition shifts to higher temperatures with increasing magnetic field, reported by Namiki $et$ $al$.[13] They proposed that this transition may be responsible for the instability of the Fermi surface $e$.$g$., charge density wave (CDW) or spin density wave (SDW) transition. In this way a dual character of 4$f$ electronic state of the Ce ions: localized and/or itinerant nature has been reported in CeOs$_{4}$Sb$_{12}$ until now.

Ultrasonic measurements are particularly suited to study via the quadrupolar response of the 4$f$ electronic ground state split by the CEF effect if the 4$f$ electrons are localized well. The elastic constants, as the quadrupolar susceptibility, measure the diagonal (Curie terms) and off-diagonal (Van Vleck terms) quadrupolar matrix elements. The quadrupolar response of the 4$f$ electronic ground state split by CEF effect causes a characteristic anomaly in the temperature dependence of the elastic constants due to the terms.[15-17] On the other hand, narrow quasiparticle bands with the possession of the enhanced effective mass cause elastic anomalies as well if the 4$f$ electrons are delocalized.[15, 18, 19] In this paper we report on the elastic properties of single-crystal CeRu$_{4}$Sb$_{12}$ and CeOs$_{4}$Sb$_{12}$ by means of ultrasonic measurements. The present results indicate that the observed elastic anomalies can not be explained well by the CEF effect in both of the systems. Alternatively, the 4$f$ electronic states of Ce ions is explained reasonably by the itinerant picture with an energy gap at the Fermi level in CeOs$_{4}$Sb$_{12}$. The preliminary reports have been published in ref. 20.

\section{\label{sec:level1}Experiment}
Single crystals of CeRu$_{4}$Sb$_{12}$ and CeOs$_{4}$Sb$_{12}$ were prepared using a molten-metal-flux growth method with Sb flux. The specimen used in our study has a size of 2.27$\times$4.14$\times$1.28 mm$^{3}$ for CeRu$_{4}$Sb$_{12}$, and 1.0$\times$0.9$\times$0.5 mm$^{3}$, respectively, but only with the crystallographic $\langle$100$\rangle$ axis for CeOs$_{4}$Sb$_{12}$. Thus, the measurements of the elastic constants $C$$_{11}$ and $C$$_{44}$ were possible for CeOs$_{4}$Sb$_{12}$ in this study. The sound velocity ($v$) was measured by an ultrasonic apparatus based on a phase comparison method in a magnetic field up to 12 T generated by a superconducting magnet. The plates of LiNbO$_{3}$ transducers were used to generate and detect the sound waves with the frequencies from 5 MHz to 30 MHz. The transducers were glued on the parallel planes of the sample by the elastic polymer Thiokol. The absolute value of the elastic constant $C$=$\rho$$v$$^{2}$ by using the density $\rho$ of the crystal was estimated explicitly for CeRu$_{4}$Sb$_{12}$ with a lattice parameter $a$ = 9.2721 $\AA$, but it was impossible for CeOs$_{4}$Sb$_{12}$ because the thickness of the sample was not enough.

\begin{figure}[h]
\begin{center}\leavevmode
\includegraphics[width=0.8\linewidth]{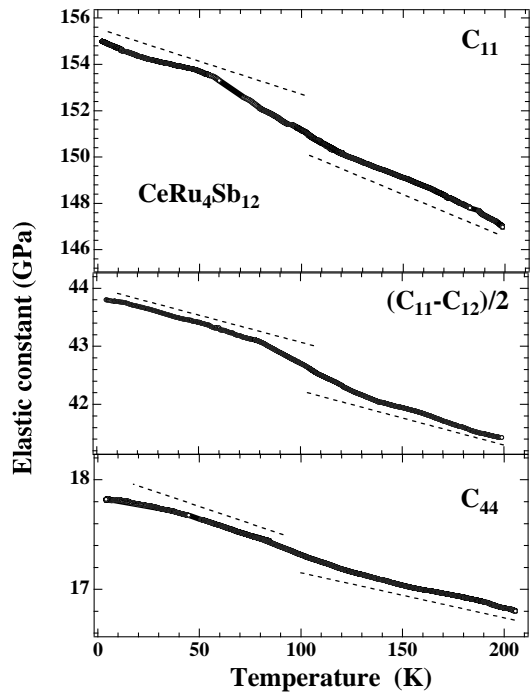}
\caption{Temperature dependence of the elastic constants $C$$_{11}$, ($C$$_{11}$-$C$$_{12}$)/2 and $C$$_{44}$ for CeRu$_{4}$Sb$_{12}$.}\label{figurename}\end{center}
\end{figure}

\begin{figure}[h]
\begin{center}\leavevmode
\includegraphics[width=0.8\linewidth]{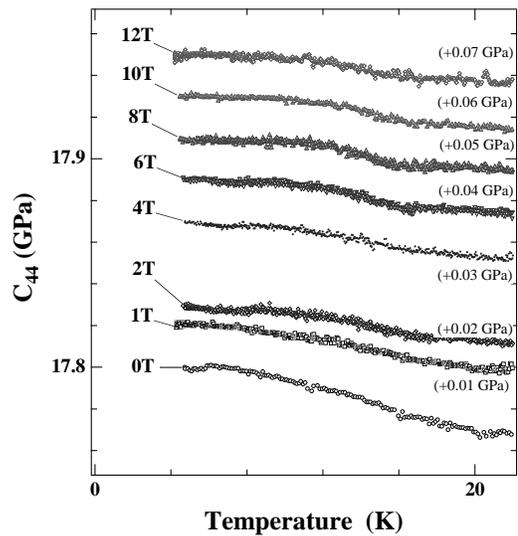}
\caption{Temperature dependence of the elastic constants $C$$_{44}$ for CeRu$_{4}$Sb$_{12}$ under selected fields along the $\langle$100$\rangle$ axis.}\label{figurename}\end{center}
\end{figure}

\begin{table}
\caption{\label{tab:tableI}The absolute values of each elastic constant, the calculated bulk modulus 
 $C$$_{B}$ =($C$$_{11}$ +2$C$$_{12}$)/3 and Poisson ratio ${\gamma}$ =$C$$_{12}$/($C$$_{11}$+$C$$_{12}$) for CeRu$_{4}$Sb$_{12}$ at both 77 and 4.2 K. }
\begin{ruledtabular}
\begin{tabular}{ccc}
\multicolumn{1}{r}{ }&\multicolumn{2}{c}{Elastic constants}\\
Mode&at 4.2 K&\mbox{at 77 K}\\
\hline
$C$$_{11}$&\mbox{155 GPa}&\mbox{152 GPa}\\
($C$$_{11}$ -$C$$_{12}$)/2&\mbox{43.8 GPa}&\mbox{43.1 GPa}\\
$C$$_{44}$&\mbox{17.8 GPa}&\mbox{17.5 GPa}\\
$C$$_{B}$ =($C$$_{11}$ +2$C$$_{12}$)/3&\mbox{96.6 GPa}&\mbox{94.5 GPa}\\
${\gamma}$ =$C$$_{12}$/($C$$_{11}$+$C$$_{12}$) &\mbox{0.303}&\mbox{0.302}\\
\end{tabular}
\end{ruledtabular}
\end{table}

\begin{figure}[h]
\begin{center}\leavevmode
\includegraphics[width=0.95\linewidth]{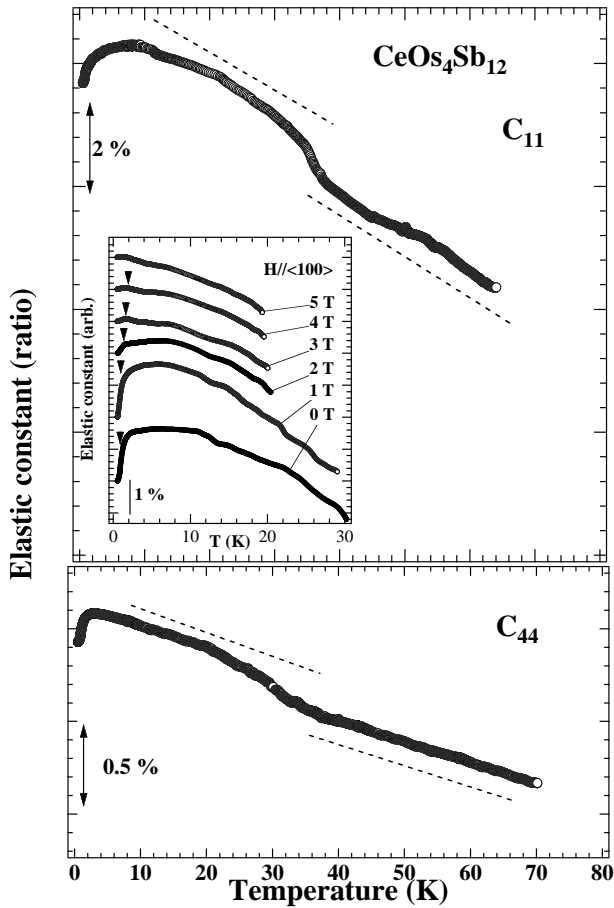}
\caption{Temperature dependence of the elastic constants $C$$_{11}$ and $C$$_{44}$ for CeOs$_{4}$Sb$_{12}$. Inset shows the temperature dependence of the $C$$_{11}$ for CeOs$_{4}$Sb$_{12}$ under selected fields along the $\langle$100$\rangle$ axis. Their offsets are shifted arbitrarily.}\label{figurename}\end{center}
\end{figure}

\begin{figure}[h]
\begin{center}\leavevmode
\includegraphics[width=0.9\linewidth]{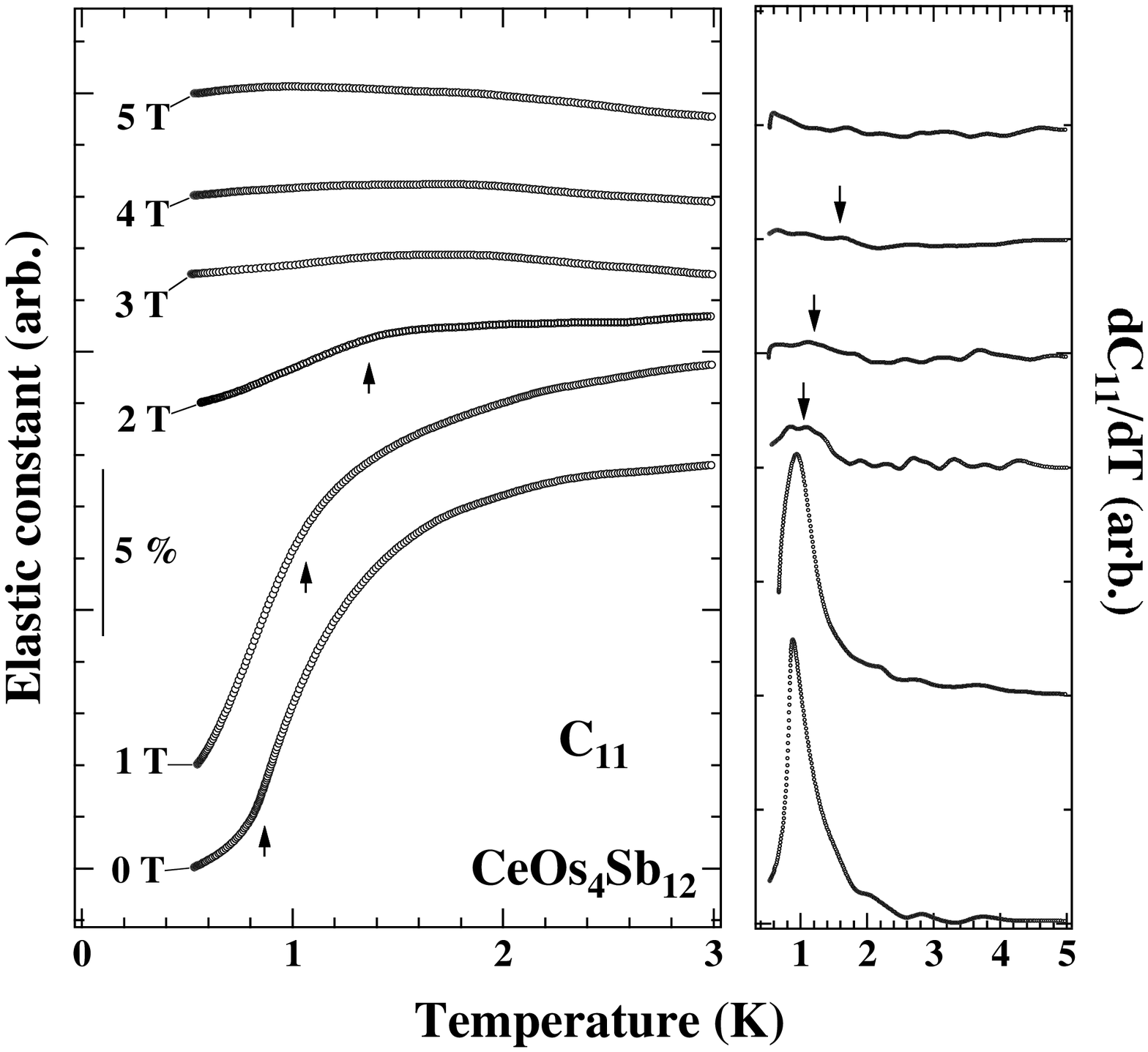}
\caption{(Left) Low temperature part of the temperature dependence of the $C$$_{11}$ under selected fields along the $\langle$100$\rangle$ axis. (Right) The corresponding d$C$$_{11}$/d$T$ curves in the low temperature region. Their offsets are shifted arbitrarily. }\label{figurename}\end{center}
\end{figure}

\begin{figure}[h]
\begin{center}\leavevmode
\includegraphics[width=0.8\linewidth]{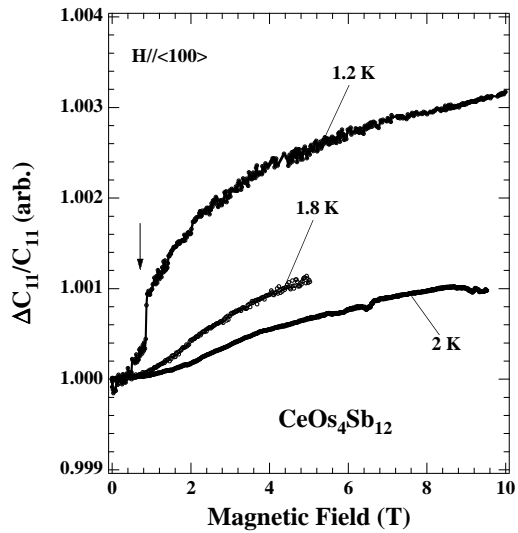}
\caption{Magnetic field dependence of the elastic constant $C$$_{11}$ at selected temperatures around the transition temperature}\label{figurename}\end{center}
\end{figure}

\begin{figure}[h]
\begin{center}\leavevmode
\includegraphics[width=0.9\linewidth]{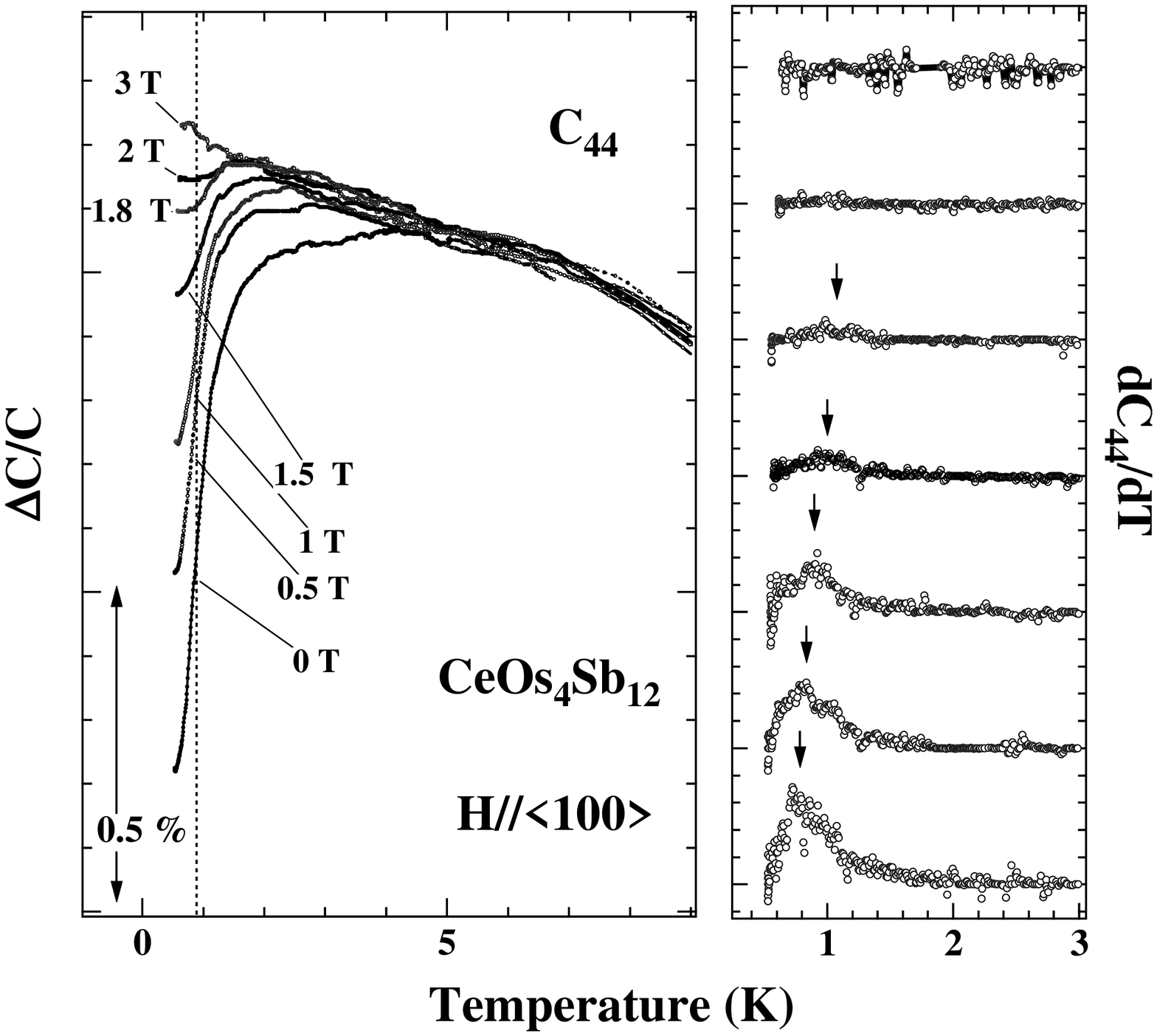}
\caption{(Left) Low temperature part of the temperature dependence of the $C$$_{44}$ under selected fields along the $\langle$100$\rangle$ axis. (Right) The corresponding d$C$$_{44}$/d$T$ curves in the low temperature region. Their offsets are shifted arbitrarily. }\label{figurename}\end{center}
\end{figure}

\begin{figure}[h]
\begin{center}\leavevmode
\includegraphics[width=0.8\linewidth]{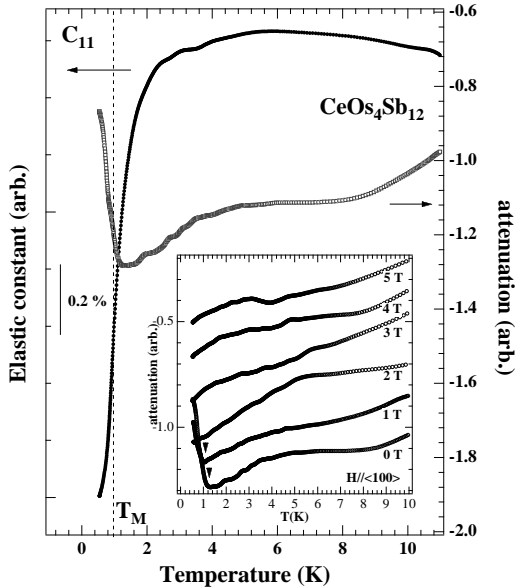}
\caption{Temperature dependence of the attenuation and the corresponding elastic constant $C$$_{11}$ in zero field. Insets shows the temperature dependence of the attenuation under selected fields along the $\langle$100$\rangle$ axis.}\label{figurename}\end{center}
\end{figure}

\begin{figure}[h]
\begin{center}\leavevmode
\includegraphics[width=0.8\linewidth]{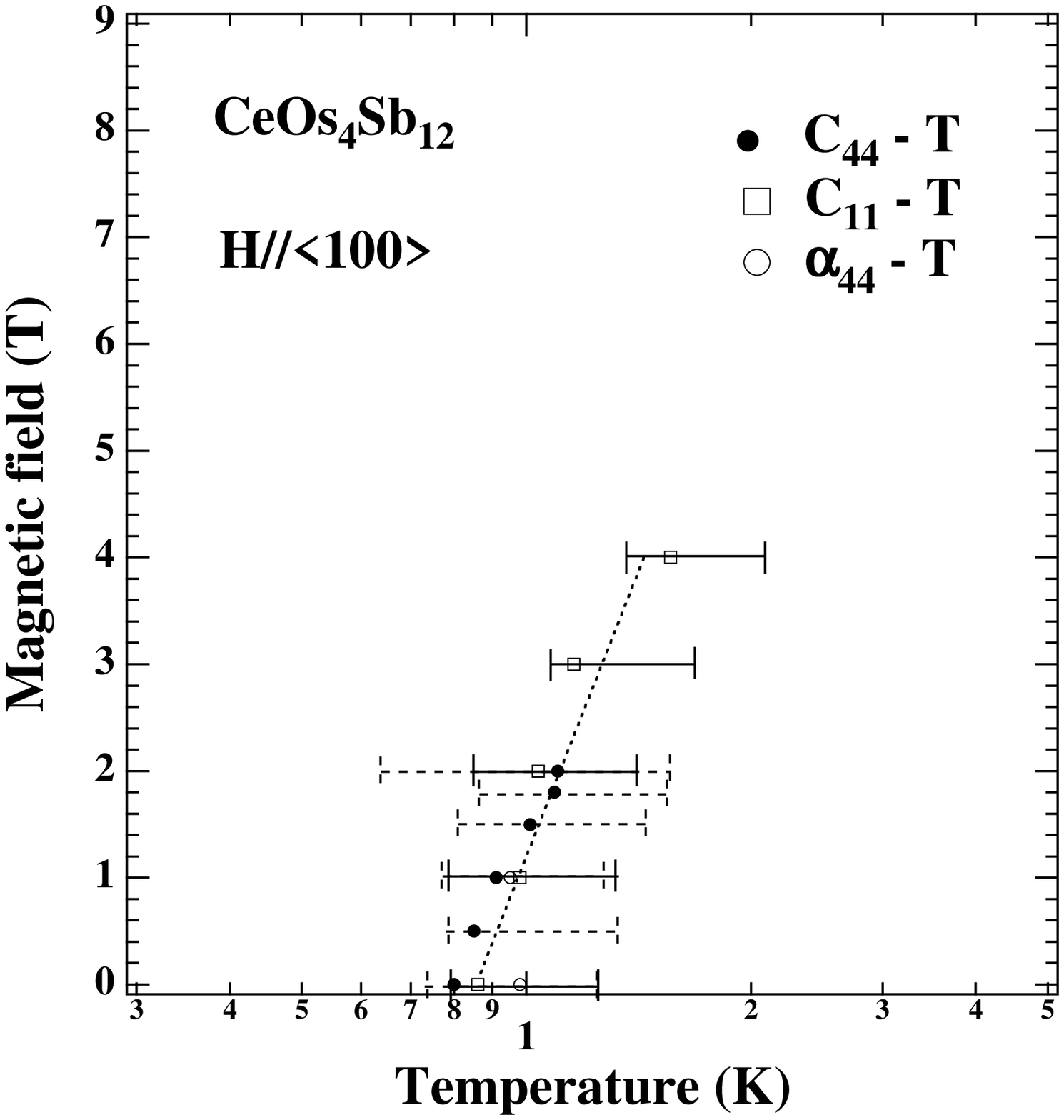}
\caption{Magnetic phase diagram for CeOs$_{4}$Sb$_{12}$ for the field along the $\langle$100$\rangle$ axis deduced from the present results. The broken line is guide to eyes. The horizontal solid and broken lines represents the error bars for $C$$_{11}$ and $C$$_{44}$, respectively. They were determine by the Full Width at Half Maximum (FWHM) of the anomaly observed around $T$$_{s}$ in the derivative of the elastic constants with respect to temperature.}\label{figurename}\end{center}
\end{figure}

\section{\label{sec:level1}Experimental results}
\subsection{\label{sec:level2}Elastic property of CeRu$_{4}$Sb$_{12}$}
Figure 1 shows elastic constants $C$$_{11}$, ($C$$_{11}$-$C$$_{12}$)/2 and $C$$_{44}$ of CeRu$_{4}$Sb$_{12}$ as a function of temperature. $C$$_{11}$ was measured by the longitudinal sound wave with frequencies of 10 - 30 MHz propagated along the $\langle$100$\rangle$ axis. ($C$$_{11}$ - $C$$_{12}$)/2 and $C$$_{44}$ were measured by the transverse sound wave with frequencies of 5 - 15 MHz propagated along the $\langle$110$\rangle$ axis with the polarization parallel to the $\langle$1-10$\rangle$ axis and propagated along the $\langle$100$\rangle$ axis with the polarization parallel to the $\langle$010$\rangle$ axis, respectively. They all increase monotonically with decreasing temperature. A slight increase was observed around 130 K in the elastic constants at which the magnetic susceptibility and electrical resistivitiy exhibit an anomalous temperature dependence.[7] This feature will be discussed in detail later. Figure 2 shows the temperature dependence of $C$$_{44}$ under selected fields along $\langle$100$\rangle$ axis. The monotonic increase shows little change even in magnetic fields up to 12 T at low temperatures within the experimental error in which the NFL behavior appears in the temperature dependence of the electrical resistivity and specific heat.[7] The absolute values of each elastic constant and calculated bulk modulus $C$$_{B}$ = ($C$$_{11}$+2$C$$_{12}$)/3, and Poisson ratio $\gamma$=$C$$_{12}$/($C$$_{11}$+$C$$_{12}$) from $C$$_{11}$ and ($C$$_{11}$-$C$$_{12}$)/2 at both 77 K and 4.2 K are listed in Table I.

\subsection{\label{sec:level2}Elastic property of CeOs$_{4}$Sb$_{12}$}
 Figure 3 shows a relative change of elastic constants $C$$_{11}$ and $C$$_{44}$ as a function of temperature. Both of the elastic constants increase monotonically with decreasing temperature. However, a pronounced softening toward low temperature was observed in $C$$_{11}$ below around 15 K, which is markedly different from that of CeRu$_{4}$Sb$_{12}$. A steep decrease was observed at 0.9 K in the elastic constants $C$$_{11}$ and $C$$_{44}$. The softening in $C$$_{44}$ toward the transition temperature $T$$_{s}$ of 0.9 K is rather slight. The degree of the softening in $C$$_{11}$ is of about 1 $\%$ down to the temperature of 0.5 K which is the lowest in this study. A hump behavior was observed around 35 K in the elastic constants $C$$_{11}$ and $C$$_{44}$. This feature will be discussed in detail later. The inset of Fig. 3 shows the temperature dependence of $C$$_{11}$ under selected fields, and the low temperature region is shown in Fig. 4, combined with the derivative of $C$$_{11}$ with respect to temperature. Their offsets are shifted arbitrarily to avoid the overlap and complication. The softening was gradually suppressed with increasing field. The middle point of the decrease shifts to higher temperatures with increasing field as indicated by the arrows. Furthermore, the steep decrease most likely to be related to the phase transition was gradually suppressed with increasing field. The softening and the steep decrease were almost undetectable above 5 T. Figure 5 shows the field dependence of $C$$_{11}$ at selected temperatures. $C$$_{11}$ increases monotonically with increasing field. However, a distinct anomaly was observed at around 1 T at the temperature of 1.2 K. This transition corresponds to the phase boundary of the ordered state mentioned above.[13-14] This phase will be discussed in detail later. Figure 6 shows the temperature dependence of $C$$_{44}$ under selected fields, combined with the derivative of $C$$_{44}$ with respect to temperature. In contrast to $C$$_{11}$, a softening toward the transition temperature of 0.9 K is slight. However, a steep decrease probably ascribed to the phase transition is observed in $C$$_{44}$ as well. Similar to $C$$_{11}$, the decrease was gradually suppressed with increasing field. Figure 7 shows a comparison between the temperature dependence of the attenuation and the corresponding elastic constant in zero field. A distinct anomaly was observed at the phase transition temperature of 0.9 K. A steep increase of the attenuation was observed below $T$$_{s}$ at which $C$$_{11}$ shows the steep decrease. The increase of the attenuation indicates increase in scattering for the propagating sound wave below 0.9 K. The anomaly shifts to lower temperatures with increasing field along the $\langle$100$\rangle$ axis as shown in the inset of Fig. 7. The magnetic phase diagram for CeOs$_{4}$Sb$_{12}$ deduced from the present results is shown in Fig. 8. The boundary goes up to the higher field side and becomes gradually obscure in the high temperature region. This nature is reminiscent of a ferromagnetic transition.

\begin{figure}[h]
\begin{center}\leavevmode
\includegraphics[width=0.8\linewidth]{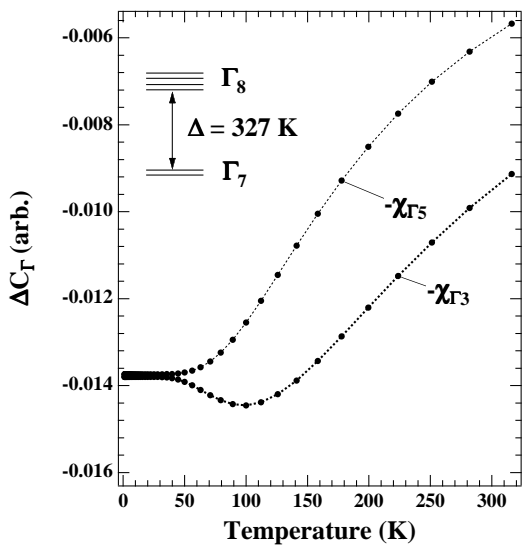}
\caption{ Theoretical results of the temperature dependence of the quadrupolar susceptibilities -$\chi$$_{\Gamma 3}$$^{(s)}$($T$) and -$\chi$$_{\Gamma 5}$$^{(s)}$($T$), belonging to the ($C$$_{11}$-$C$$_{12}$)/2 and $C$$_{44}$, respectively, for CeOs$_{4}$Sb$_{12}$ using eq. (2). }\label{figurename}\end{center}
\end{figure}

\begin{figure}[h]
\begin{center}\leavevmode
\includegraphics[width=0.8\linewidth]{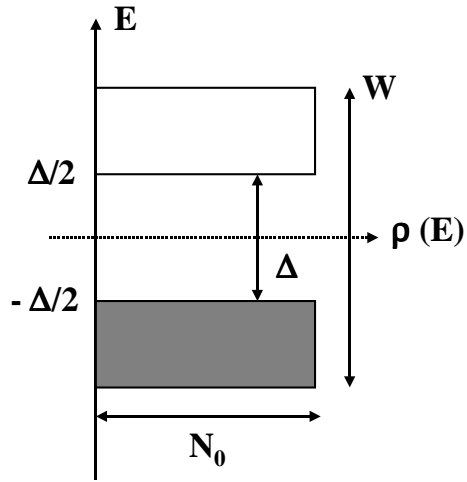}
\caption{Schematic representation of the densities of the 4$f$-electronic state for CeOs$_{4}$Sb$_{12}$ symmetrically lying above and below $E$$_{F}$.}\label{figurename}\end{center}
\end{figure}

\begin{figure}[h]
\begin{center}\leavevmode
\includegraphics[width=0.8\linewidth]{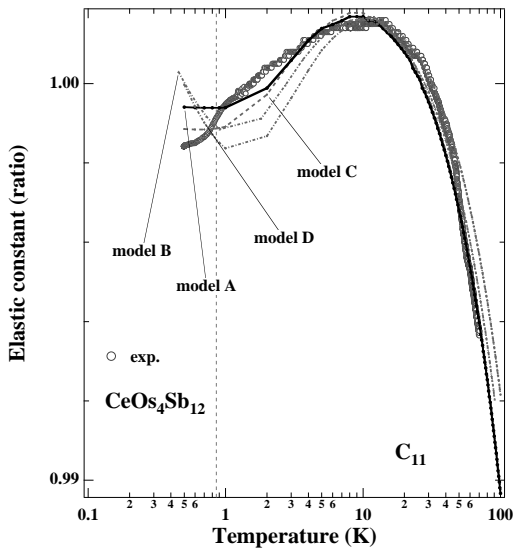}
\caption{Theoretical results of the temperature dependence of the elastic constant $C$$_{11}$ based on the deformation potential using eq. (5).}\label{figurename}\end{center}
\end{figure}

\begin{table}
\caption{\label{tab:tableI}Obtained parameters deduced from the present results as shown in Fig. 11.}
\begin{ruledtabular}
\begin{tabular}{ccc}
\multicolumn{1}{r}{ }\\
model&W&$\Delta$\\
\hline
A&\mbox{30 K}&\mbox{5 K}\\
B&\mbox{30 K}&\mbox{6 K}\\
C&\mbox{30 K}&\mbox{6 K}\\
D&\mbox{29 K}&\mbox{5.5 K}\\
\end{tabular}
\end{ruledtabular}
\end{table}

\section{\label{sec:level1}Discussion}
Firstly, we will discuss the obtained results of CeRu$_{4}$Sb$_{12}$. The slight increase observed around 130 K in the temperature dependence of $C$$_{11}$, ($C$$_{11}$-$C$$_{12}$)/2 and $C$$_{44}$ seems to be related the anomalous behavior in the temperature dependence of the resistivity and magnetic susceptibility. The magnetic susceptibility begins to deviate significantly from the Curie-Weiss law expected in a trivalent Ce ion state below 150 K and exhibits a broad peak around 100 K followed by an upturn below 50 K.[7] This characteristic behavior is seen in intermediate-valence (IV) compounds. Furthermore, a slope of the temperature dependence of the electrical resistivity is changed below around 150 K. A rapid decrease is observed below 80 K. [7] The present results indicate that the slight increase is related to the onset of the deviation of the magnetic susceptibility from the Curie-Weiss law. This may suggest that the valence state of Ce in CeRu$_{4}$Sb$_{12}$ begins to be unstable below around 150 K. In the case, the charge fluctuation of the Ce ions usually has the strong influence on the bulk modulus $C$$_{B}$ since it is related closely to the change of the total volume as seen in a representative of the mixed valence system SmB$_{6}$.[21] The slight increase observed around 150 K, thus seems to be ascribable to the change of the valence. However, it would be difficult to conclude that the slight increase is due to the instability of the electronic configuration of the Ce ion at this stage, because the magnitude of the anomaly in $C$$_{B}$ for CeRu$_{4}$Sb$_{12}$ is much smaller than that for SmB$_{6}$ in which the mixed valence state is formed. Nevertheless, again, this interpretation is consistent with that the valence state of Ce in CeRu$_{4}$Sb$_{12}$ becomes unstable at low temperatures suggested by the magnetic susceptibility and the lattice constant measurements.[7, 8] Most probably, a strong hybridization between the conduction electrons and the 4$f$-electronic state plays a crucial role at the temperature in this system. Few influence of the NFL state on the elastic constants is recognized in this study. This might indicate that the origin is attributed to an irrelevant magnetic fluctuation, but not the relevant charge one to the elastic strain.

Let us move on the discussion of CeOs$_{4}$Sb$_{12}$. First, we mention the CEF level scheme of the Ce in CeOs$_{4}$Sb$_{12}$. The elastic softening toward low temperature, observed in well-localized 4$f$ electron systems, can be usually understood as the quadrupolar response of the system to external strain. This effect has its origin in the modulation of the CEF by the strain. If one neglects the inter-site interaction between quarrupolar moments, the elastic softening toward low temperature can be analyzed with the following formula (1).[22-24]

\begin{equation}
C_{\Gamma}(T)= -Ng_{\Gamma}^2\chi_{\Gamma}^{(s)} (T)
\end{equation}

Here, $N$ and g$_{\Gamma}$ are the number of Ce ions in unit volume and the coupling constant between the quadrupolar moment and the relevant elastic strain, respectively. $\Gamma$ represents the irreducible representation: $\Gamma$$_{3}$ and $\Gamma$$_{5}$ for a cubic system. $\chi$$_{\Gamma}$$^{(s)}$($T$) denotes the quadrupolar susceptibility for the 4$f$ electronic state in the cubic CEF potential, which can be written as follows, 

\begin{eqnarray}
\chi_{\Gamma}^{(s)}(T)=\sum_{\text{ik}}\frac{exp(-E_{ik}^{(0)}/k_{B}T)}{Z}\nonumber\\
\times
\left(\frac{1}{k_{B}T}\vert \langle ik| O_{\Gamma}|ik\rangle \vert ^2-2\sum_{\text{jl}}\frac{\vert \langle ik| O_{\Gamma}|jl\rangle \vert ^2}{E_{i}-E_{j}}\right)
\end{eqnarray}

where $|ik\rangle $ represents the $k$-th eigenfunction of the $i$-th CEF level. If the ground state is degenerate with respect to a quadrupolar moment $O$$_{\Gamma}$, a softening in the corresponding elastic constant is expected to occur due to the non-zero Curie term. Figure 9 shows the calculated results of a relative change of the elastic constant, $\Delta $$C$$_{\Gamma}$ for ($C$$_{11}$-$C$$_{12}$)/2=-$N$g$_{\Gamma 3}$$\chi$$_{\Gamma 3}$$^{(s)}$($T$) and $C$$_{44}$=-$N$g$_{\Gamma 5}$$\chi$$_{\Gamma 5}$$^{(s)}$($T$) based on the proposed CEF level scheme in CeOs$_{4}$Sb$_{12}$.[9] The calculated quadrupolar susceptibility $\chi$$_{\Gamma 3}$$^{(s)}$($T$) and $\chi$$_{\Gamma 5}$$^{(s)}$($T$) are belonging to the elastic constants both ($C$$_{11}$-$C$$_{12}$)/2 and $C$$_{44}$, respectively. Unfortunately it is impossible to compare quantitatively the present results to the theoretical ones based on the CEF effect since ($C$$_{11}$-$C$$_{12}$)/2 could not be measured in this study. Nevertheless, if the CEF effect is dominant at low temperature i.e., a well-localized picture can be adapted to CeOs$_{4}$Sb$_{12}$, a pronounced softening is expected in both ($C$$_{11}$-$C$$_{12}$)/2 and $C$$_{44}$. In particular, a minimum is expected in the temperature dependence of ($C$$_{11}$-$C$$_{12}$)/2 around $\Delta$/2= 165 K, where $\Delta$ denotes the splitting energy of CEF effect as shown in Fig. 9. $C$$_{11}$ consists of a linear combination of ($C$$_{11}$-$C$$_{12}$)/2 and bulk modulus $C$$_{B}$. Thus, $C$$_{11}$ is qualitatively equivalent to ($C$$_{11}$-$C$$_{12}$)/2 if the CeOs$_{4}$Sb$_{12}$ is the well-localized 4$f$-system. If so, it is quite hard to account for the obtained results: a softening below 15 K in $C$$_{11}$ only by the CEF effect in CeOs$_{4}$Sb$_{12}$. We need to postulate a different way to explain this softening. Here, we argue the origin of the observed elastic softening below 15 K in $C$$_{11}$ as follows. In the case of an electronic band with a large cyclotron effective mass, a coupling between the conduction electrons and the relevant elastic strains associated with a sound wave can be significant, and causes anomaly in the elastic constant as mentioned in Introduction.[18-19] This essential coupling may be described by
\begin{eqnarray}
E_{k}= E_{k}^{0}+g_{k \Gamma}\varepsilon_{\Gamma}+(h_{k \Gamma})^{2}\varepsilon_{\Gamma}^{2}/\Delta_{k}
\end{eqnarray}

where, $E$$_{k}$, $g$$_{k}$ and $h$$_{k}$ denote the energy of the upper or lower quasiparticle band, the first and the second deformation potential coupling constants, respectively.  $\Delta$$_{k}$ denotes the band gap at the Fermi level. Then, the free energy of the conduction electrons is described as follows,

\begin{eqnarray}
F_{el}=n{\zeta}-k_{B}T\sum_{\text{k}}ln(1+exp(\frac{\zeta-E_{k}}{k_{B}T}))
\end{eqnarray}

Here, $n$ denotes the number of the conduction electrons in the quasi-particle band and $\zeta $ is the chemical potential. From the second derivative of the free energy with respect to the elastic strain $\varepsilon$$_{\Gamma}$, one obtains an analogous expression for the elastic constants as in the case of magnetoelastic interaction with Fermi-Dirac distribution function $f$$_{k}$={1+exp(($\zeta$-$E$$_{k}$)$/$k$_{B}$$T$)}$^{-1}$ as follows,

\begin{eqnarray}
C_{\Gamma}(T)= \frac{\partial^{2}F}{\partial\varepsilon_{\Gamma}^{2}}=C_{\Gamma}^{0}+\sum_{\text{k}}\frac{\partial^{2}E_{k}}{\partial\varepsilon_{\Gamma}^{2}}f_{k}\nonumber\\
-\frac{1}{k_{B}T}\sum_{\text{k}}(\frac{\partial E_{k}}{\partial\varepsilon_{\Gamma}})^{2}f_{k}(1-f_{k}) \nonumber\\
+\frac{1}{k_{B}T}\frac{(\sum_{\text{k}}\frac{\partial E_{k}}{\partial\varepsilon_{\Gamma}}f_{k}(1-f_{k}))^{2}}{\sum_{\text{k}}f_{k}(1-f_{k})}
\end{eqnarray}

The second term and the both of third and fourth terms of eq. (5) represent the Van Vleck and the Curie term, respectively. $C$$_{\Gamma}$$^{0}$ in eq. (5) is the background elastic constant without the contribution of the $f$ electrons. Variation in $C$$_{\Gamma}$$^{0}$ originates mainly from anharmonic effects of the crystal. This formalism is very similar to that of the strain susceptibility of the localized 4$f$-electron system without the Fermi distribution function $f$$_{k}$. Here, the conservation law of the total number of electrons is employed in the quasi-particle band. For simplicity, the simple quasi-particle band model is assumed, in which the dispersion of band energy is neglected and the Fermi level $\varepsilon$$_{F}$ is located at the middle of energy gap $\Delta$. In addition, g$_{k}$ and $\Delta$$_{k}$ are independent of the wave vector $k$. Furthermore, the rectangular density of states $N$$_{0}$ for the upper and lower bands with the same band width $W$ were introduced. The present model is illustrated in Fig. 10. It is noted that a shift of the chemical potential is not expected with changing the temperature for this symmetric two-band model.

We have analyzed the elastic softening with eq. (5) under the above-mentioned conditions. The fitting results are shown by the dotted lines in Fig. 11. Here, the background elastic constant was determined so that the high temperature part followed the obtained results. The obtained parameter was summarized in table II. The estimated value of the band gap $\Delta$ is in good agreement with that determined by the electrical resistivity measurement.[11] This value is, however quite small compared to that determined by the optical conductivity measurement.[25] In the latter case, it is almost of ten times in magnitude. This inconsistency may be ascribed to the complicated band structure in the vicinity of Fermi level $\varepsilon$$_{F}$, and that the band gap has the strong temperature dependence as generally expected in conventional Dense Kondo semiconductors and semimetals. The multiple formation of the energy gap may be realized in CeOs$_{4}$Sb$_{12}$, which is developing by the change of temperature. 

Secondary, we discuss the phase transition at $T$$_{s}$ of 0.9 K. As mentioned in Introduction, this is expected to be due to the instability of the Fermi surface, such as charge density wave (CDW) or spin density wave (SDW) transition because of the extremely tiny entropy release of 0.02Rln2 at this temperature.[13-14] A steep decrease was observed in the elastic constants at $T$$_{s}$, which is quite different from that observed in a conventional magnetic ordering due to the well-localized 4$f$ electronic state.[17, 23] A clear upturn, $i$.$e$., an abrupt elastic hardening is observed in that case such as NdFe$_{4}$P$_{12}$ which exhibits a ferromagnetic transition at 1.9 K with full release of the magnetic entropy of $R$ln4 expected by the quartet ground state.[2, 26] This experimental fact may also indicate that the phase transition at $T$$_{s}$ is due more to the conduction electron state than the localized 4$f$-electronic state. Interestingly, this transition shifts to higher temperatures with increasing field, which is consistent with the results of the specific heat measurement. This behavior let us conjecture that of a quadrupolar ordering as seen in CeB$_{6}$ and TmTe, or a ferromagnetic transition.[27-29] This model, however cannot explain the extremely tiny entropy release at $T$$_{s}$ as mentioned above. Recently, a ferromagnetic ordering with the tiny entropy release was reported in some other filled skutterudite compounds such as Pr$_{1-x}$La$_{x}$Fe$_{4}$P$_{12}$ being for $x$=0.05 and 0.15, and SmFe$_{4}$P$_{12}$ in which the extremely heavy fermion system is realized.[30, 31] We suggested that their systems are close to the quantum critical point (QCP) just followed disappearance of a magnetic or quadrupolar ordering phase.[32] It is pointed out that the instability of Fermi surface arisen from a strong hybridization plays a crucial role for this ferromagnetic transition. Again, the present results let us conjecture that such a peculiar magnetic transition may occur at $T$$_{s}$ of 0.9 K in CeOs$_{4}$Sb$_{12}$. The origin of the phase transition in CeOs$_{4}$Sb$_{12}$ is left to the future studies.

Thirdly, we discuss the selective connection between the elastic softening and the Fermi surface from the viewpoint of their symmetry in CeOs$_{4}$Sb$_{12}$. According to the de Haas- van Alphen effect measurement and the band calculation of the reference material LaOs$_{4}$Sb$_{12}$, the Fermi surface consists of the closed hole surface derived from the 47th band centered at the $\Gamma$ point and the closed and multiply connected hole surfaces from 48th band.[33, 34] Especially, the multiply connected hole surface is relevant to the heavy electrons through the correlation effect, whose main parts are centered at the N point.[33] The band structure calculations in LaOs$_{4}$Sb$_{12}$ indicate that the Fermi surface of the band is missing in the direction of $\langle$100$\rangle$ and $\langle$111$\rangle$, and their equivalent. It can therefore be presumed that this multiply connected hole surfaces has the strong deformation coupling with the relevant elastic strain induced by the sound wave. If we assume a charge density located at the N point, the character for them is decomposed into a direct sum of the irreducible representations $\Gamma$$_{1}$, $\Gamma$$_{3}$ and $\Gamma$$_{4}$. Consequently, there can be charge-fluctuation modes with $\Gamma$$_{3}$ symmetry, which possibly couple to the elastic strains $\varepsilon$$_{u}$=(2$\varepsilon$$_{zz}$-$\varepsilon$$_{xx}$-$\varepsilon$$_{yy}$) or $\varepsilon$$_{v}$=($\varepsilon$$_{xx}$-$\varepsilon$$_{yy}$) with $\Gamma$$_{3}$ symmetry of the soft elastic mode $i$.$e$., ($C$$_{11}$-$C$$_{12}$)/2 mode and $C$$_{11}$ as well. The lack of $\Gamma$$_{5}$ representation expects the absence of the significant elastic softening around $T$$_{s}$ in the C$_{44}$ mode associated with the elastic strains $\varepsilon$$_{yz}$, $\varepsilon$$_{zx}$, $\varepsilon$$_{xy}$ of $\Gamma$$_{5}$ symmetry. This interpretation can explain reasonably the obtained results.

In addition, a slight change was observed in the temperature dependence of $C$$_{11}$ around 30 K in CeOs$_{4}$Sb$_{12}$. It would be difficult to conclude at present whether this anomaly is intrinsic or not. The reproducibility of the date was confirmed in the present study. However, no remarkable anomaly around this temperature has been observed in the other physical properties. If this is intrinsic this may be due to crystalline electric field (CEF) effect proposed by Bauer $et$ $al$,.[11] Actually, the proposed CEF level scheme is expected to cause a change of the slope in the temperature dependence of $C$$_{11}$ and $C$$_{44}$ around 35 K. Another scenario is the thermally activated rattling motions of an off-center rare-earth atom in a cage of Sb-icosahedron, recently suggested by Goto $et$ $al$,.[35, 36] The increase behavior of the elastic constant with decreasing temperature lets us conjecture that of PrOs$_{4}$Sb$_{12}$. Very recently, we found a same behavior in the temperature dependence of the elastic constants of SmOs$_{4}$Sb$_{12}$ [37]. These experimental facts suggest that the rattling motion of an off-center rare-earth atom is common feature in REOs$_{4}$Sb$_{12}$. A relatively large lattice parameter or/and a relatively large size of the cage of Sb-icosahedron in REOs$_{4}$Sb$_{12}$ may yield a condition for the rattling motion. This is left to future studies, and further experiments including the sound attenuation measurement with a large size of the single-crystal are strongly requisite.

Finally, we would like to make a comment on the common elastic feature in the Filled Skutterudite compounds that show semiconductor-like behavior in their electrical resistivity. PrRu$_{4}$P$_{12}$ and SmRu$_{4}$P$_{12}$ exhibit a metal-insulator transition at 62.4 K and 16 K, respectively.[38, 39] The electrical resistivity increases rapidly below the transition with decreasing temperature in the both compounds. Except for the phase transition the similar behavior is seen in that of CeOs$_{4}$Sb$_{12}$. Interestingly, the characteristic softening toward low temperature was observed also in the temperature dependence of elastic constants of PrRu$_{4}$P$_{12}$ and SmRu$_{4}$P$_{12}$.[40, 41] The common origin may cause the characteristic softening in these compounds. As we will report in the separated articles, a clear softening was observed in the temperature dependence of the elastic constants(C$_{11}$-C$_{12}$)/2 and C$_{44}$ below about 2 K in SmRu$_{4}$P$_{12}$. However, it is quite difficult to explain them reasonably by the CEF effect.[41] In PrRu$_{4}$P$_{12}$ a clear excitation of the CEF level scheme was observed by the inelastic neutron scattering measurement.[42] The ground state of Pr ion split by the CEF effect expects such a characteristic softening in the temperature dependence. In these systems, it would be difficult to determine the primary origin to cause the elastic softening, $i$.$e$., being ascribable to the deformation potential or/and the CEF effect only by results of ultrasonic measurements. Again, we believe that the coupling between the elastic strain $\varepsilon$$_{u}$ or $\varepsilon$$_{v}$ and the heavy conduction electrons whose band is centered at the N point, plays a clue role in CeOs$_{4}$Sb$_{12}$ in causing the elastic anomaly in $C$$_{11}$. On the other hand, the characteristic elastic softening was not observed in CeRu$_{4}$Sb$_{12}$ which exhibits a metallic behavior at low temperatures, exactly speaking non-Fermi liquid behavior. The different manner in the temperature dependence of the elastic constants between CeOs$_{4}$Sb$_{12}$ and isostructural CeRu$_{4}$Sb$_{12}$ may provide an evidence of the gap formation in CeOs$_{4}$Sb$_{12}$.The different electronic band structure in the vicinity of the Fermi level  would govern the observed elastic anomalies. Interestingly, the similar feature is seen in the Dense Kondo compounds CeNiSn and isostructural CePdSn as well.[19]

\section{\label{sec:level1}Summary}
In this paper we have presented the elastic properties of the non-Fermi liquid metal CeRu$_{4}$Sb$_{12}$ and the dense Kondo semiconductor CeOs$_{4}$Sb$_{12}$. We have performed ultrasonic measurements to measure the temperature dependence and the field dependence of the elastic constants $C$$_{11}$, ($C$$_{11}$-$C$$_{12}$)/2 and $C$$_{44}$ for CeRu$_{4}$Sb$_{12}$, and $C$$_{11}$ and $C$$_{44}$ for CeOs$_{4}$Sb$_{12}$, respectively. From our experiments the following conclusions have been obtained. A slight increase was observed around 130 K in the elastic constants of CeRu$_{4}$Sb$_{12}$, where the magnetic susceptibility and the electrical resistivitiy exhibit an anomalous temperature dependence. The instability of the valence state in the Ce ion may cause the elastic anomaly around 130 K. In CeOs$_{4}$Sb$_{12}$, on the other hand, a pronounced softening in the longitudinal $C$$_{11}$ toward the $T$$_{s}$ was observed. This can be explained reasonably by the coupling between the elastic strain $\varepsilon$$_{u}$ or $\varepsilon$$_{v}$ and the heavy conduction electrons whose band is centered at the N point. The energy gap $\Delta$ and bandwidth $W$ were estimated to be about 5 to 6 K and 30 K, respectively. It is indicated, from this work, that the quesiparticles formed by the hybridization between the conduction electron band and the 4$f$ localized electrons play a crucial role in the low temperature properties of CeOs$_{4}$Sb$_{12}$. Furthermore, a steep decrease was observed in the elastic constants around $T$$_{s}$. Although the origin of this phase is not clear at this stage, this elastic behavior is likely to be common in the other skutterudite compounds where a week magnetic ordering with a small amount of entropy release occurs at low temperatures. This behavior lets us conjecture the magnetic ordering originated in the itinerant heavy quasiparticles. For a further discussion, it is necessary to make clear the magnetic structure in the ordered phase in CeOs$_{4}$Sb$_{12}$.

\begin{acknowledgments}
The authors thank M. Nakamura for support in the experiments. The measurements have been performed in the Cryogenic Division of the Center for Instrumental Analysis, Iwate University. This work was supported by a Grant-in-Aid for Science Research Priority Area "Skutterudite" (No. 15072202) of the Minister of Education, Culture, Sports, Science, and Technology of Japan.
\end{acknowledgments}

\end{document}